\documentclass[ twocolumn, preprint]{aastex631}
  
\usepackage{xcolor}
\usepackage{amsmath,amstext}
\usepackage[T1]{fontenc}
\usepackage{apjfonts} 
\usepackage{natbib}
\usepackage{microtype}
\usepackage{verbatim}
\usepackage{hyperref}
\hypersetup{
    colorlinks=true,
    linkcolor=blue,
    filecolor=magenta,      
    urlcolor=cyan,
}

\newcommand{\hst}{\textit{HST}}

\newcommand{\jwst}{\textit{JWST}}
\newcommand{\threedhst}{\hbox{3D-HST}}

\newcommand{\zg}{\hbox{$z_{grism}$}}

\newcommand{\lsig}{\hbox{$\log(\Sigma_1)$}}
\newcommand{\Lsig}{\hbox{$\log(\Sigma_1)$}}
\newcommand{\Lmass}{\hbox{$\log(M / M_\odot)$}}
\newcommand{\lmass}{\hbox{$\log(M / M_\odot)$}}

\newcommand{\lssfr}{\hbox{$\log(sSFR)$}}
\newcommand{\Lssfr}{\hbox{$\log(sSFR)$}}

\newcommand{\psf}{\hbox{$P_{Q}$}}
\newcommand{\pq}{\hbox{$P_{Q}$}}

\newcommand{\Panel}[1]{\colorbox{black!52}{\textcolor{white}{#1}}}
\newcommand{\editone}[1]{\textcolor{black}{#1}}
\definecolor{aggiemaroon}{HTML}{500000}

\begin{document}

\title{\large \bf CLEAR: The Morphological Evolution of Galaxies in the Green Valley}

\author[0000-0001-8489-2349]{Vicente Estrada-Carpenter}
\affiliation{Department of Physics and Astronomy, Texas A\&M University, College
Station, TX, 77843-4242 USA}
\affiliation{George P.\ and Cynthia Woods Mitchell Institute for
 Fundamental Physics and Astronomy, Texas A\&M University, College
 Station, TX, 77843-4242 USA}
 \affiliation{Department of Astronomy \& Physics, Saint Mary's University, 923 Robie Street, Halifax, NS, B3H 3C3, Canada}

\author[0000-0001-7503-8482]{Casey Papovich}
\affiliation{Department of Physics and Astronomy, Texas A\&M University, College
Station, TX, 77843-4242 USA}
\affiliation{George P.\ and Cynthia Woods Mitchell Institute for
 Fundamental Physics and Astronomy, Texas A\&M University, College
 Station, TX, 77843-4242 USA}

\author[0000-0003-1665-2073]{Ivelina Momcheva}
\affil{Space Telescope Science Institute, 3700 San Martin Drive,
  Baltimore, MD, 21218 USA}
\affil{Max-Planck-Institut für Astronomie, Königstuhl 17, D-69117 Heidelberg, Germany}

\author[0000-0003-2680-005X]{Gabriel Brammer}
\affil{Cosmic Dawn Centre, University of Copenhagen, Blegdamsvej 17, 2100 Copenhagen, Denmark}

\author[0000-0002-6386-7299]{Raymond C. Simons}
\affil{Space Telescope Science Institute, 3700 San Martin Drive,
  Baltimore, MD, 21218 USA}


\author[0000-0001-7151-009X]{Nikko J. Cleri}
\affiliation{Department of Physics and Astronomy, Texas A\&M University, College Station, TX, 77843-4242 USA}
\affiliation{George P.\ and Cynthia Woods Mitchell Institute for Fundamental Physics and Astronomy, Texas A\&M University, College Station, TX, 77843-4242 USA}



\author{Mauro Giavalisco}
\affil{Astronomy Department, University of Massachusetts,
Amherst, MA, 01003 USA}


\author[0000-0002-7547-3385]{Jasleen Matharu}
\affiliation{Department of Physics and Astronomy, Texas A\&M University, College
Station, TX, 77843-4242 USA}
\affiliation{George P.\ and Cynthia Woods Mitchell Institute for
 Fundamental Physics and Astronomy, Texas A\&M University, College
 Station, TX, 77843-4242 USA}

\author[0000-0002-1410-0470]{Jonathan R. Trump}
\affil{Department of Physics, University of Connecticut, Storrs, CT 06269, USA}

\author[0000-0001-6065-7483]{Benjamin Weiner}
\affil{MMT/Steward Observatory, 933 N. Cherry St., University of Arizona, Tucson,
AZ 85721, USA}

\author[0000-0001-7673-2257]{Zhiyuan Ji}
\affiliation{Steward Observatory, University of Arizona, 933 N. Cherry Avenue, Tucson, AZ 85721, USA}

\begin{abstract}
Quiescent galaxies having more compact morphologies than star-forming galaxies has been a consistent result in the field of galaxy evolution. What is not clear is at what point this divergence happens, i.e. when do quiescent galaxies become compact, and how big of a role does the \editone{progenitor effect} play in this result? Here we aim to model the morphological and star-formation histories of high redshift (0.8 $<$ z $<$ 1.65) massive galaxies (\Lmass\ $>$ 10.2) with stellar population fits using HST/WFC3 G102 and G141 grism spectra plus photometry from the CLEAR (CANDELS Lyman-alpha Emission at Reionization) survey, constraining the star-formation histories for a sample of $\sim$ 400 massive galaxies using flexible star-formation histories. We develop a novel approach to classifying galaxies by their formation activity in a way that highlights the green valley population, by modeling the specific star-formation rate distributions as a function of redshift and deriving the probability that a galaxy is quiescent (\pq). Using \pq\ and our flexible star-formation histories we outline the evolutionary paths of our galaxies in relation to stellar mass, Sersic index, $R_{eff}$, and stellar mass surface density. We find that galaxies show no appreciable stellar mass growth after entering the green valley (a net decrease of 4$\%$) while their stellar mass surface densities increase by $\sim$ 0.25 dex. Therefore galaxies are becoming compact during the green valley and this is due to increases in Sersic index and decreases in $R_{eff}$.
\end{abstract}

\section{Introduction}

Recent work in the field of galaxy evolution has begun to outline the formation and quenching pathways of massive quiescent galaxies at high redshifts \citep[e.g.,][]{beli19, estr19, krie19, estr20, sues20, tacc21, ji22}. A major goal of this field of study is to understand how the star formation, chemical enrichment, morphological, and quenching histories of galaxies evolve with redshift. Because these evolutionary processes happened at a more rapid pace at high redshift \citep{tran07, eise08, blak09, papo10, thom10} we can better constrain these formation history properties with greater certainty \citep{estr19}. This is an area of study that has benefited significantly from modern telescopes, instruments, and methods as these galaxies become difficult to study at high redshift because of lower luminosity due to the domination of older stellar populations, smaller sample sizes as the fraction of quiescent galaxies shrinks at higher redshifts \citep{muzz13, tomc14, kawi17}, and a high density of telluric emission lines contaminating the rest-frame optical in ground-based telescopes \citep{sull12}.

Here, we focus on the star-formation history (SFH) of galaxies in this study. SFHs estimate galaxies' formation activity (e.g., star-formation rate - SFR) and predict when their stellar masses were formed. Simple approaches, such as color-color diagrams, have been used to approximate a galaxy's formation. The UVJ diagram \citep{wuyt07, will09, whit11} is one such method that has been widely adopted due to its ability to separate star-forming and quiescent galaxies. More advanced methods model the SFH as a function of time; these methods can be as simple as using single stellar population models (effectively a delta function), to parametric models such as $\tau$ (exponentially dropping SFR) and delayed-$\tau$ (a linear rise followed by an exponential drop in SFR) models. Recently, the development of easy-to-use, flexible, and non-parametric star-formation history codes \citep[for example][]{leja19, iyer17} have allowed for the derivation of more natural formation histories. As we combine these more realistic formation histories with various timescales and morphological properties, we can begin to trace how specific properties of galaxies evolve.

One question we can address with realistic SFHs is when do quiescent galaxies become compact. When compared to star-forming galaxies, quiescent galaxies as a population have more compact morphologies \citep{vanw14}. The implication is that quiescent galaxies should have gone through some morphological transition in their past, although some other effects such as the \editone{progenitor effect} \citep{ji22a} can also contribute to the apparent morphological evolution \citep{lill16, ji22a}. \editone{Progenitor effect} is caused by galaxies that quench at lower redshifts having more extended morphologies than galaxies that quenched at higher redshifts, and therefore the overall quiescent population will have mean morphologies that are more extended at lower redshifts.  This effect is compounded due to the higher percentage of quiescent galaxies at lower redshifts \citep{kawi17}. \editone{Progenitor effect is similar to progenitor bias \citep{vand99}, where younger/recently quenched galaxies get added to quiescent population at lower redshifts, but adds that as these galaxies quenched at lower redshifts they will also have more extended morphologies.} 

\editone{Similar works at lower redshifts have studied the evolution of morphology (stellar mass surface density - \cite{fang13, wu20, guo21}, and bulge/disk - \cite{brem18, kim18, lope18, quil22}) as galaxies cross the green valley using various techniques to identify the green valley (colors, specific SFR, morphology, and absorption lines). They broadly find that galaxies become more bulge-dominated as they cross the green valley. The two main mechanisms used to explain this evolution are bugle growth and disk fading \citep{math20}. Bulge growth would require stellar mass to be added to the bulge during the green valley, while disk fading requires no change to the morphology and is the result of a younger disk component fading leaving the older bulge component unchanged. Bulge growth during the green valley being the dominant factor would mean that galaxies continue to morphologically evolve as they transition, while disk fading being dominant would mean that a galaxies morphology is set when it enters the green valley.} The question we aim to answer is, does the transition from extended star-forming galaxy to compact quiescent galaxy happen while the galaxy is in its star-forming phase (disk fading), does it happen while it is transitioning to a quiescent galaxy (bulge growth), i.e., crossing the green valley, or is this relationship totally dominated by the \editone{progenitor effect}?

In this work we develop a novel approach to identifying transitioning galaxies by deriving the probability that a galaxy is quiescent, \psf\ using the shape of the specific star-formation rate (sSFR) distribution with a large sample ($\sim$ 400) of massive galaxies (\Lmass\ $>$ 10.2) at 0.8 $<$ z $<$ 1.65.  We compare  the morphological properties of star-forming galaxies, quiescent galaxies, and ``transitioning'' galaxies, to  determine when they become compact. In Section \ref{sec_data} we describe our data and sample selection. In Section \ref{sec_met} we outline our fitting methodology and development of \psf. In Section \ref{sec_res} we discuss our modeling of \psf. In Section \ref{sec_dis} we describe the implications of the evolutionary tracks we see for quiescent and green valley galaxies. Finally, in Section \ref{sec_con} we summarize our results. Throughout this work, we assume a cosmology with $\Omega_{m,0}=0.3$, $\Omega_{\lambda,0}=0.7$, and $H_0=70$ km s$^{-1}$. 

\section{Data \label{sec_data}}
\begin{figure*}[t]
\epsscale{1.15}
\plotone{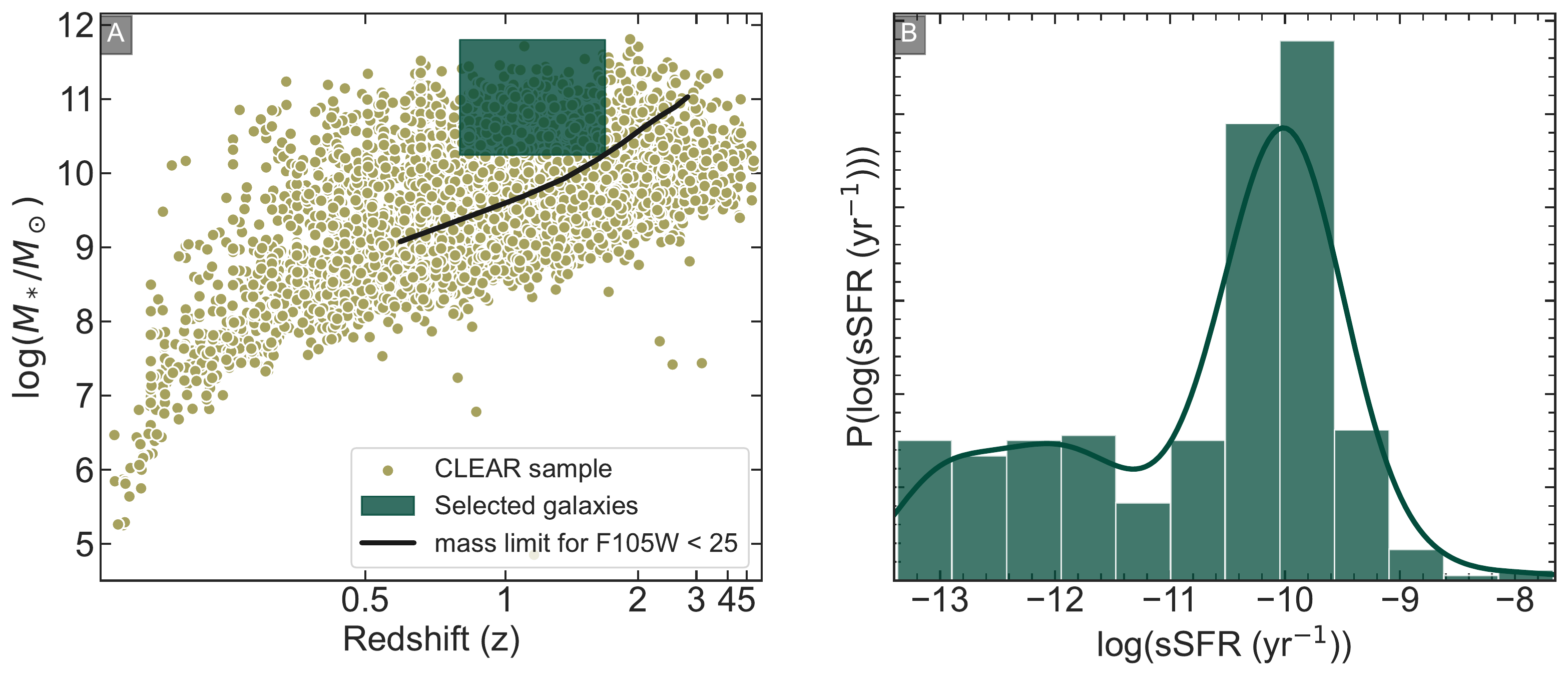}
\caption{Panel \Panel{A} shows the CLEAR sample (yellow) with stellar masses and redshifts measured by Eazy-py. The black curve in \Panel{A} corresponds to the mass required for a single stellar population formed at z = 5 to be detected with AB magnitude of F105W < 25 mag (CLEAR magnitude limit). The volume-limited sample used in this paper is shown in the green box with (\Lmass\ $>$ 10.25 and 0.8 $<$ \zg\ $<$ 1.65) and was chosen to maximize the amount of low \Lssfr\ ($<$ -11) galaxies. Panel \Panel{B} shows the corresponding \Lssfr\ distribution which is complete to \Lmass\ $>$ 10.25 at all redshifts. The volume-limited sample (defined by this green-shaded region in \Panel{A}) ensures quiescent galaxies are properly represented at each redshift. 
\label{fig_sample}}
\end{figure*} 

For this work, we use data from the CLEAR (CANDELS Lyman$-\alpha$ Emission at Reionization) survey which includes \hst\ WFC3/G102 grism spectra, \hst\ WFC3/G141 grism spectra, and photometry (see \cite{estr19, estr20, papo22, simo21, simo23}. Our parent sample selection consists of massive galaxies (\Lmass\ $>$ 9.8), using masses derived from \texttt{eazy-py} \citep{bram08, eazy-py} within a redshift range of 0.8 $< z <$ 2.8 using z$_{grism}$ measurements from \texttt{grizli} \citep{grizli} and z$_{phot}$ from \texttt{eazy-py}, details in \citet{simo23}. \editone{We set our redshift lower limit to z = 0.8 as this places H$\beta$ and OIII at the bluest end of the HST/WFC3 G102 spectra. At z < 0.8 it becomes more difficult to measure the redshifts from the grism data and subsequently stellar populations are less constrained.} We use both redshift measurements as the z$_{grism}$ measurements accurately recover redshifts for emission-line galaxies while the z$_{phot}$ measurements are more reliable for quiescent galaxies \citep[see][]{estr20}. Finally, we examine the 1D spectra for every exposure by eye to look for any residual contamination, where we either remove it (if possible) or exclude the galaxy if it is too contaminated. We are then left with a sample of 1390 galaxies. These galaxies were all then fit for their stellar populations using the methods described in Section \ref{sec_met}. We then remove any galaxies from our sample which have X-ray detections likely from AGN \citep{xue16,luo17}. Our parent sample selection is shown in Figure \ref{fig_sample}.

We then apply several cuts to the parent sample for our stellar population fits. As we are interested in the sSFR distribution we are careful to select a sample with a complete representation of high and low sSFRs, therefore we use a volume-limited, mass-complete sample. To do this we found the minimum stellar mass needed for a detection of Y$_{F105W} = 25$ AB mag (magnitude limit used for extraction in the CLEAR dataset) by modeling a single stellar population (SSP) formed at z $=$ 5, shown in Figure \ref{fig_sample} Panel \Panel{A} as a black curve. \editone{The SSP models a galaxy that formed and quenched at high redshift and therefore provides a measurement of the Y$_{F105W} = 25$ AB mag of an older/fainter quiescent galaxy, providing the lower limit where we could reasonably recover quiescent galaxies.} Our final selection window is shown as the green shaded region. This region was selected as it maximizes our sample of low sSFR galaxies (\Lssfr\ $<$ -11) resulting in a selection window of \Lmass\ $>$ 10.25 and 0.8 $<$ \zg\ $<$ 1.65, with a sample of 384 galaxies. Figure \ref{fig_sample} Panel \Panel{B} shows the \Lssfr\ distribution of our sample. Here we see that the use of the volume-limited sample generates a bimodal \Lssfr\ distribution in the final sample.
 
\begin{figure*}[t]
\epsscale{1.}
\plotone{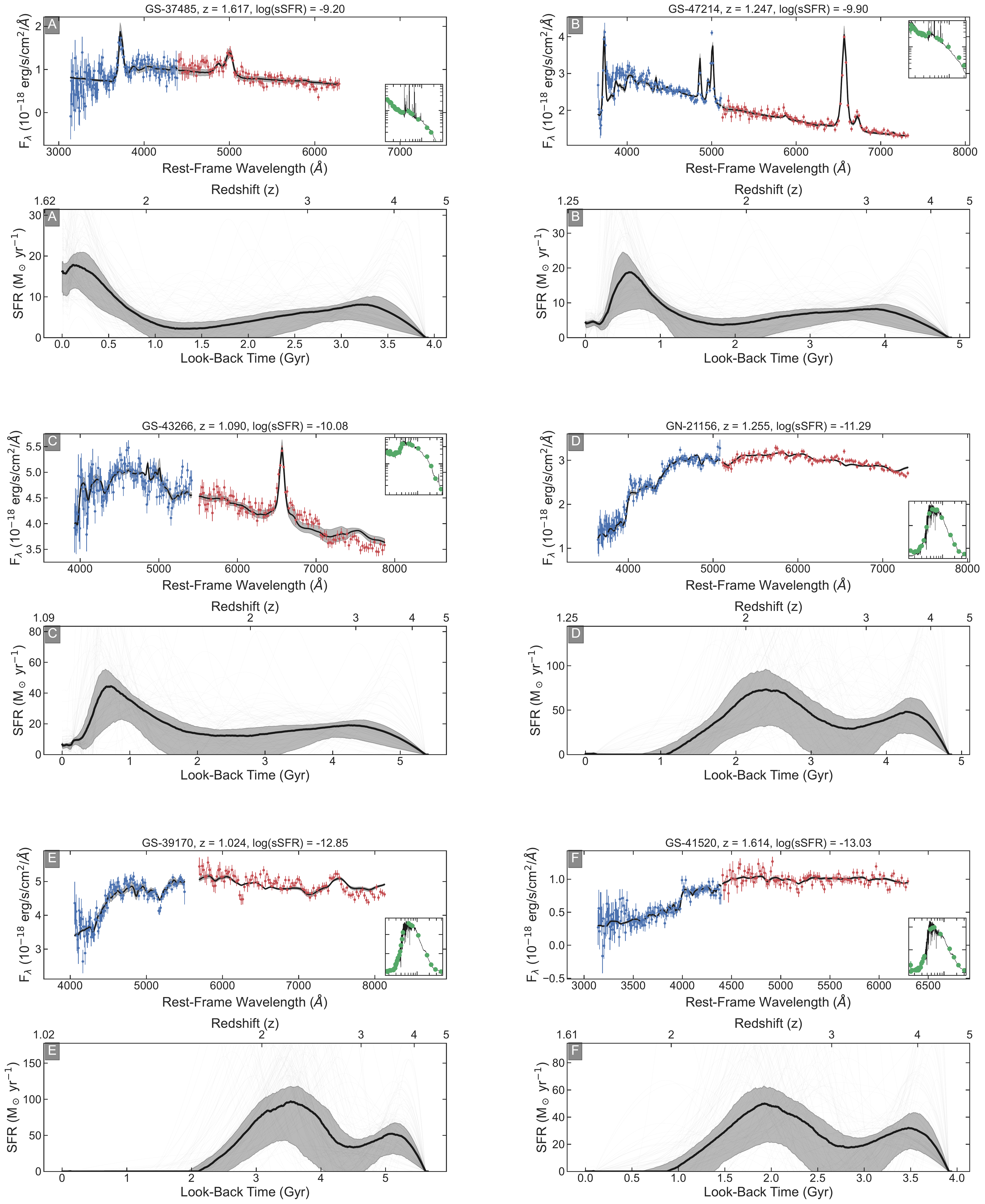}
\caption{Example spectra and posterior SEDs (black) from our stellar population fits with \hst\ WFC3/G102 spectrum in blue, \hst\ WFC3/G141 spectrum in red, and photometry in green (inset panel, shown in log(F$_\lambda$) for Panels \Panel{A} - \Panel{C} and F$_\lambda$ for Panels \Panel{D} - \Panel{F}), along with posterior SFHs (50th percentile in black with the inner 68 percentile shaded grey and individual draws from the posterior shown as thin black lines) on the bottom panel plotted against look-back time (with redshift shown on top). The panels from \Panel{A} to \Panel{F} progress to lower \Lssfr\ values, with features of the spectra resembling a more mature stellar population and the SFHs showing a reduction in stellar mass formation. 
\label{fig_spec}}
\end{figure*} 

\section{Method \label{sec_met}}

To derive the stellar populations of our galaxies, we use the forward modeling approach from \cite{estr19}, the flexible SFHs of Dense Basis from \cite{iyer17}, SED models utilizing the Flexible Stellar Population Synthesis (FSPS) models \citep{conr10}, with MILeS and BaSeL libraries and assuming a Kroupa initial mass function \cite{krou01}, and sampling with the nested sampling algorithm Dynesty \citep{spea19}. Utilizing the full capacity of the CLEAR data set allows us to simultaneously fit \hst\ WFC3/G102 grism spectra, \hst\ WFC3/G141 grism spectra, and photometry which covers rest-frame UV to IR ($\sim$ 31 bands in GOODS-South and $\sim$ 20 bands in GOODS-North)

Our approach to fitting spectra and photometry simultaneously is to assume that the flux calibration of the photometry is truth, but allow the grism spectra to be scaled along with the models. This is due to different approaches being used in extracting the photometry and grism spectra, therefore one should assume that there is some flux offset between the two. The grism spectra are fit in 2D at every exposure and orientation. Additionally, we are fitting the emission lines using the Cloudy integration in FSPS \cite{ferl13,byle17}.

The stellar population parameters we fit are:
\begin{itemize}
    \item log(sSFR$_{dense basis}$ (yr$^{-1}$)) - used to generate the SFHs, using a flat prior over a range of (-14,-7)
    \item t$_{x}$ (Gyr)- This includes t$_{25}$, t$_{50}$, t$_{75}$ which are the timescales at which the galaxy formed 25$\%$, 50$\%$, and 75$\%$ of its stellar mass, using a Dirichlet prior, used to generate the SFHs
    \item Stellar metallicity (log(Z/Z$_\odot$)) - flat prior over a range of (-2.25, 0.25)
    \item Nebular metallicity (log(Z/Z$_\odot$)) - flat prior over a range of (-2.25, 0.25), independent of the stellar metallicity
    \item Ionization parameter (log(U)) - flat prior over a range of (-3.5,-2)
    \item Dust (Av) - assumed the Calzetti dust law \citep{calz00}, using a log prior over a range of (0,2)
    \item redshift (z) - using a Gaussian prior with the mean set to the \texttt{grizli} or \texttt{eazy-py} (see \ref{sec_data}) derived redshift and a $\sigma$ value of $\sim$ 0.04 z
    \item Nuisance parameters - a set of parameters to apply a slight tilt to the SEDs to account for possible contamination over/under subtraction
\end{itemize}

There are also several parameters that we derive post-fitting, 
\begin{itemize}
    \item t$_{90}$ (Gyr) - Timescale at which the galaxy formed 90$\%$ of its stellar mass, derived from posterior SFH
    \item t$_q$(Gyr) - Quenching timescale defined as t$_{50}$ - t$_{90}$, derived from posterior SFH
    \item log(sSFR (yr$^{-1}$)) - The sSFR averaged over the last 100 Myr, derived from posterior SFH, note that this is different from the sSFR used to derive the SFH in our SED fitting though they only differ by -0.02$_{-0.22}^{+0.06}$ dex
    \item SFR (M$_\odot$/yr) - The SFR averaged over the last 100 Myr, derived from posterior SFH
    \item SFR$_{peak}$ (M$_\odot$/yr) - The peak SFR, derived from posterior SFH
    \item $\log(M/M_\odot)$ - Stellar mass, derived from the optimized scaling coefficient used in our SED fitting
    \item \editone{$\log(\Sigma_1 (M_\odot/kpc^{-2}))$ - Stellar-mass surface density within 1 kpc, derived using our measurements of stellar mass and the Sersic profiles from \cite{vanw14}. These are calculated by integrating the Sersic profile out to 1 kpc and then dividing it by the Sersic profile integrated out to infinity (see equations \ref{eq1}, \ref{eq2})} 
\end{itemize}

Figure \ref{fig_spec} shows the \hst\ WFC3/G102 grism spectra (blue), \hst\ WFC3/G141 grism spectra (red), and photometry (green-inset panel- shown in log(F$_\lambda$) in Panels \Panel{A}\ - \Panel{C} and F$_\lambda$ in Panels \Panel{D}\ - \Panel{F}) for six different galaxies with posterior SEDs (black). The spectra represent galaxies with differing star-formation activity (e.g. star-forming, transitioning, quiescent). We see here that our fitting approach models the grism spectra + photometry well, and does so for all types of galaxies. We also include the posterior SFH for each galaxy in the lower sub-panel with the SFH in black and the 68$\%$ confidence interval in grey (with individual draws from the posteriors in thin black lines). The SFHs are plotted in look-back time, therefore t = 0 here represents when the galaxy was observed and time progresses towards the Big Bang (redshift values are shown on the top of the plots). Here we see the many different shapes that our flexible SFHs can produce.

\subsection{The Quiescent Galaxy Probability, \psf}
Most methods to classify galaxy formation activity end up using a bimodal approach, either classifying galaxies as star-forming or quiescent. This approach is adequate for the general population of galaxies as most galaxies show traditional traits of either star-forming or quiescent galaxies. Where these approaches fail is in classifying galaxies that have ambiguous traits. For most classification schemes, galaxies that transitioning from star-forming to quiescence sit at the separating line, and are often treated as contamination in either sample. Likewise, when these schemes are used to try and identify green valley galaxies \editone{\cite{fang18, verg18,math20,angt21, noir22}} it is likely that the green valley population will be contaminated with both quiescent and star-forming galaxies. 

Our approach here is to assign to each galaxy its probability of being quiescent, \psf. Most galaxies have traditional quiescent or star-forming features. These galaxies will therefore lie at the extremes of our probability distribution as there is little ambiguity to their star-formation activity status. The advantage of \pq\ is in identifying galaxies that do not easily fall into traditional quiescent or star-forming roles, galaxies that (depending on the selection criteria) could fall into either classification and therefore this method excels in selecting galaxies that are crossing the green valley.

To estimate \psf\ we use the shape of the \Lssfr\ distribution. Panel \Panel{B}\ in Figure \ref{fig_sample} provides an example of the sSFR distribution. Here we can see that the measured sSFR distribution is bimodal, and therefore we can treat the sSFR distribution as the summation of the star-forming and quiescent galaxy sSFR distribution. The true shape of the sSFR distribution is likely not bimodal and is instead probably a distribution with a long tail \citep{feld17}, but the bimodality of the measured sSFR distribution comes from the inability of SED fitting codes to measure extremely low SFRs. These codes fail to recover SFRs for galaxies with low or no star formation, and the expectation values will be scattered to higher values of SFR. This upward scatter is code dependent, and in testing our fitting code (by fitting mock spectra with SFRs = 0 $M_\odot$/yr) we found that they scattered up to \Lssfr\ values with a mean of -12 with a variance of $\sim$ 0.6 dex. While this upward scattering is problematic, the values of the scattered points are still squarely in the quiescent region of the sSFR distribution. These upward scattered points create the  ``bump'' seen in the quiescent region of the $\log sSFR$ bimodal distribution that we identify as the quiescent peak.  \citet{feld17} discuss this effect in more detail, but our conclusions are broadly similar. In the end, these upward scattered points will not be a problem in our modeling as our results are not affected by these points. 

\begin{figure*}[th]
\epsscale{1.15}
\plotone{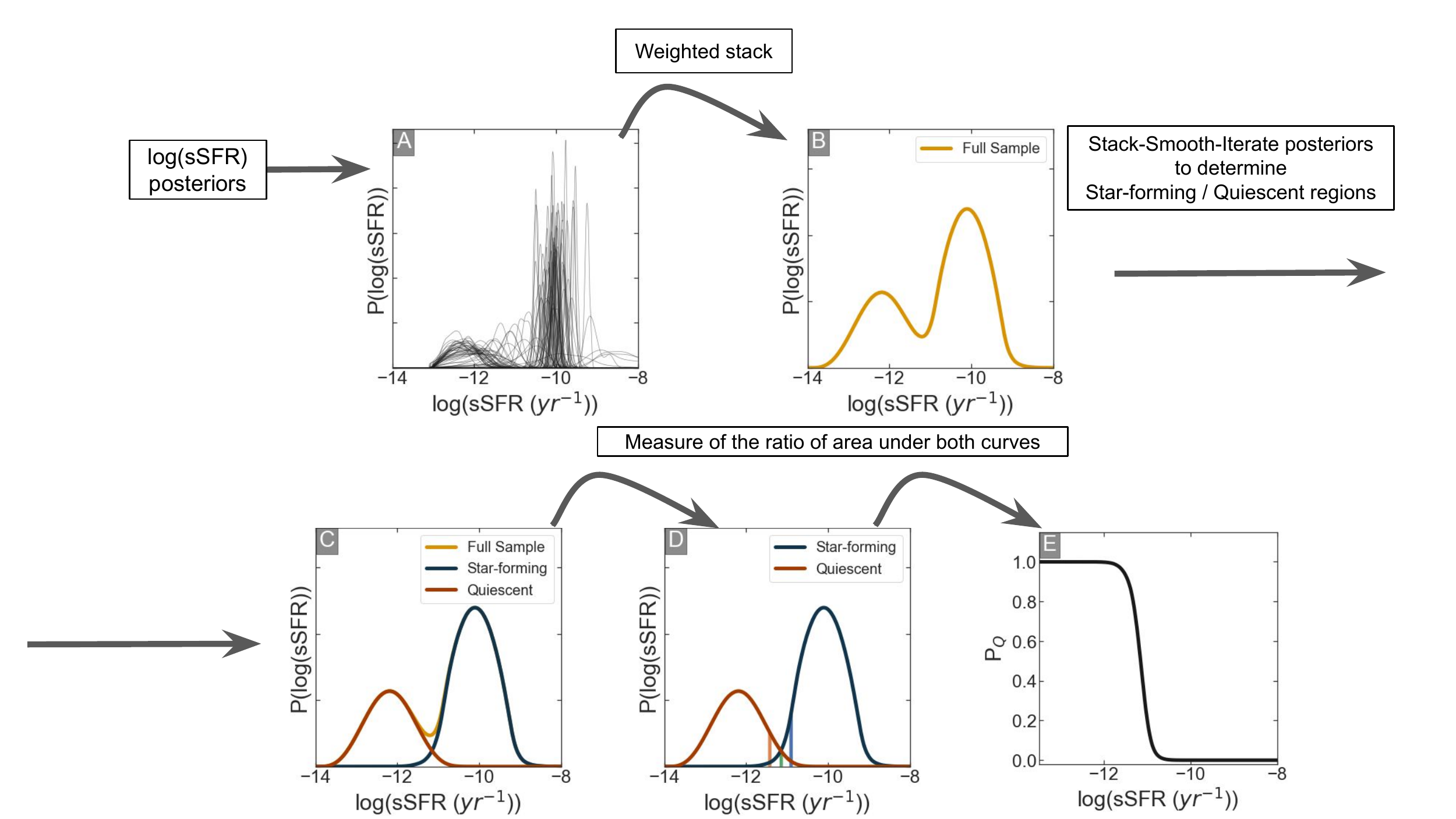}
\caption{An outline of how \psf\ is derived. First, we gather the \Lssfr\ posteriors for galaxies in the redshift bin we are examining, then we stack using the weighted stacking method outlined in \cite{estr19}. We then use the stack-smooth-iterate algorithm defined in \cite{estr19} to derive the star-forming and quiescent regions. Panel \Panel{C} shows the resulting fit.  \psf\ is then derived by measuring the area under the quiescent distribution and comparing that to the stacked distribution to derive its contribution. Panel \Panel{D} is similar to \Panel{C}, but the vertical lines show the 10$\%$ (blue), 50$\%$ (green), and 90$\%$ (red) \psf\ regions. The resulting measurement of \psf\ versus \Lssfr\ shown in \Panel{E} shows that \Lssfr\ it is dominated by high and low \pq\ values with a short transitional period. \pq\ can be thought of as a way to classify galaxies as a function of \Lssfr\ which emphasizes the green valley.
\label{fig_outline}}
\end{figure*} 

\subsection{Modeling the sSFR Distribution and Measuring \psf}

Figure \ref{fig_outline} outlines how we model the \Lssfr\ distribution and derive \pq. Step 1 is to select galaxies within a certain window of look-back time \editone{from the present}. Because the \Lssfr\ distribution may evolve with redshift we derive \psf\ in bins of redshift. Here we chose  bins of redshift that span 1.5 Gyrs as it provides samples that are just large enough to provide a "smooth" measurement. We then apply a weighted stack \citep[see][]{estr19} to the \Lssfr\ posteriors, generating the \Lssfr\ distribution for that sample. Step 2 is to then model the star-forming and quiescent distributions ($D_{sf}$, $D_{Q}$ respectively) using the stack-smooth-iterate methodology developed in \cite{estr19}. The purpose of stack-smooth-iterate is to take a set of posteriors and derive a parent distribution. This works by first multiplying the posteriors by a prior (in the first iteration the prior is flat) then stacking the posteriors using a weighted stacking method \citep{estr19}, then smoothing this distribution to remove any noisy peaks. The resulting distribution is then set as our prior and we repeat the process until the distribution converges (this is accomplished in $\sim$20 iterations). To derive  $D_{sf}$ and $D_{Q}$ we split the sample at various sSFR values, derive the parent samples, we then use a KS test to determine which $D_{sf}$ and $D_{Q}$ recreate the overall sample best, as can be seen in Figure \ref{fig_outline} Panel \Panel{C} where the overall sample (yellow) is overlayed by the $D_{sf}$ (blue) and $D_{Q}$ (red). Step 3) with our two regions, we can now measure \psf(\Lssfr) using the following function.
\begin{eqnarray}
    x = \log{(sSFR)},\\
    \epsilon = 0.01,\\
    P_{Q}(x) \equiv \frac{\int_{x - \epsilon}^{x + \epsilon} D_{Q}(x') dx'}
    {\int_{x - \epsilon}^{x + \epsilon} D_{sf}(x') dx' + 
    \int_{x - \epsilon}^{x + \epsilon} D_{Q}(x') dx'}
\end{eqnarray}
Panel \Panel{D} shows the measured values of \pq\ $=$ 0.1, 0.5, 0.9 as blue, green, red vertical lines, respectively. The \psf\ values as a function of \Lssfr\ are shown in the Panel \Panel{E} of Figure \ref{fig_outline}. Much of the \lssfr\ space is dominated by values of 1 or 0 (quiescent or star-forming) with a  short transitional period. These results mirror what is seen in most studies, that the transition from star-forming to quiescent is short-lived with \editone{ green valley crossing timescales of $\sim$ 0.5 Gyrs (\cite{blan22}, simulated galaxies) to $\sim$ 1 Gyr (\cite{noir22}, observed galaxies)}.

At its core \psf\ is a remapping of \Lssfr\ which accounts for the redshift evolution of the star-forming main sequence and highlights galaxies in transition between the star-forming and quiescent populations. \pq\ does not place galaxies into different groups but instead provides a continuous classification scheme by estimating the likelihood that galaxies belong to each group.  With that said, there is a value in defining galaxies into their respective groups, i.e., quiescent, transitioning, and star-forming, because these allow us a natural way of comparing the properties of each galaxy population.  Therefore, for the remainder of this work, we define quiescent galaxies as galaxies with a very high probability of being quiescent (\psf\ $>$ 0.99), star-forming galaxies as galaxies with a very low probability of being quiescent (\psf\ $<$ 0.01), and galaxies which are crossing the green valley as galaxies with 0.01 $<$ \psf\ $<$ 0.99. \editone{We note that \pq\ is sample dependent as a higher mass sample will have a different sSFR distribution and would therefore change the \pq\ values slightly. \pq\ acts more like a scale than a probability as quiescent and star-forming galaxies sit at the extremes of the scale (\pq\ $\sim$ 1 and \pq\ $\sim$ 0 respectively) and anything in between is seen as having an ambiguous classification and therefore we classify it as transitioning. A galaxy with \pq\ = 0.7 could be thought of as having spectral features 70$\%$ like a quiescent galaxy and 30$\%$ like a star-forming galaxy.} 

\section{Results \label{sec_res}}
\begin{figure}[th]
\epsscale{1.2}
\plotone{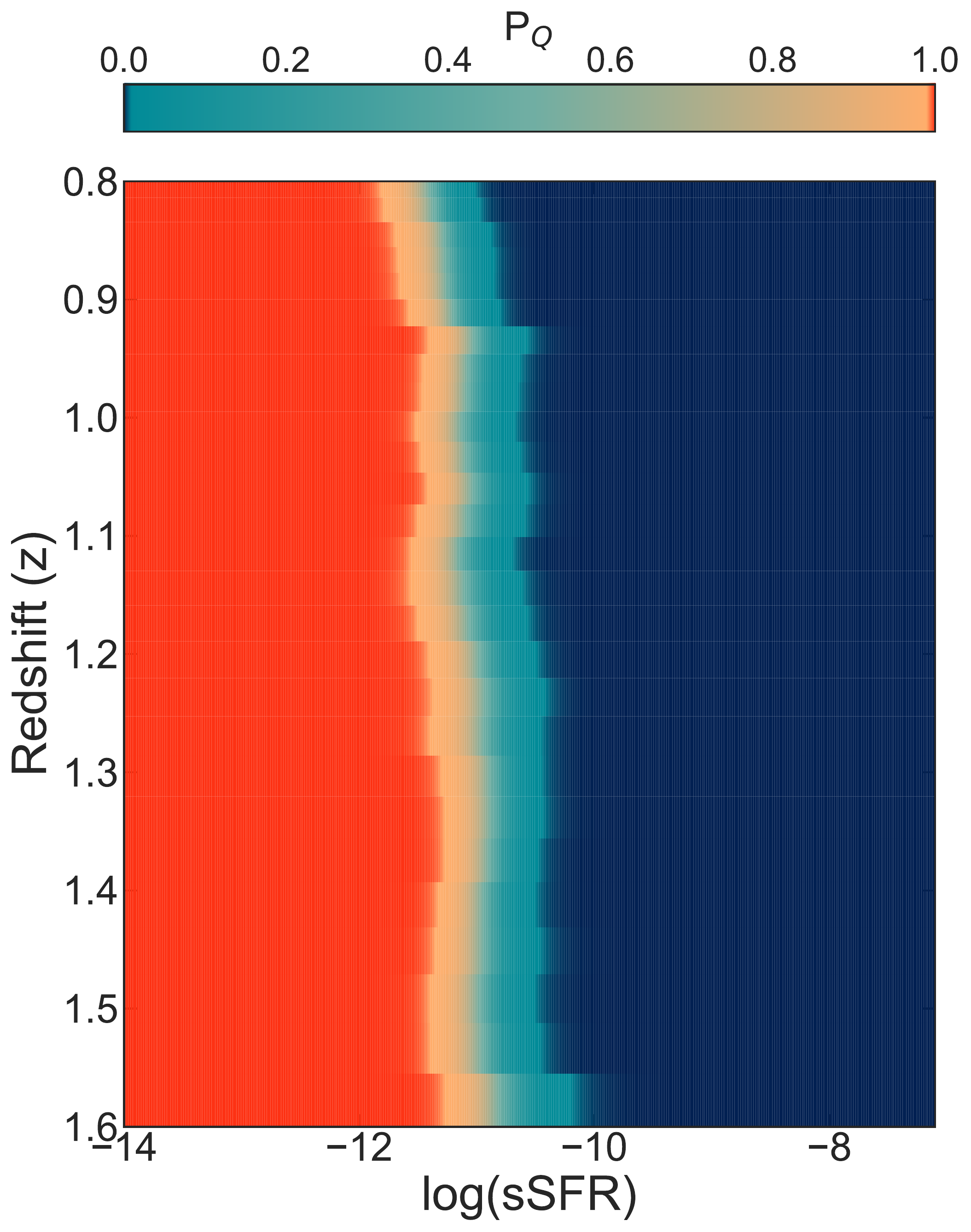}
\caption{The evolution of \pq\ as a function of redshift. The majority of the sample is dominated by star-forming and quiescent galaxies with the transitional region occupying a smaller space. The width of the transitional region does not evolve as a function of redshift, while as a whole evolving to lower \Lssfr\ at lower redshifts, a similar behavior to what is seen in the evolution of the star-forming main sequence \citep{whit12} and the green valley from \cite{tacc21}.
\label{fig_Psf}}
\end{figure} 

\subsection{\psf\ Measurement  }
Our measurement of \psf\ is shown in Figure \ref{fig_Psf}. Our method (which is outlined in Figure \ref{fig_outline}) is applied as a function of redshift. In Figure \ref{fig_Psf}, the color scaling is defined such that quiescent galaxies are shown in red, star-forming galaxies are shown in blue, and galaxies in transition are shown with a greenish hue that scales from blue to red with $P_Q$. As in Figure \ref{fig_outline}, the distribution of \Lssfr\ is dominated by \pq\ $\approx$ 0 or $\approx$ 1 with only a brief transition period 0 $<$ \psf\ $<$ 1.  That is, the vast majority of galaxies are either star-forming ($\sim$60$\%$ of the population) or quiescent ($\sim$30$\%$ of the population), with only a few in transition ($\sim$10$\%$ of the population).   Figure \ref{fig_Psf} shows that the sSFRs of transitional galaxies evolve with redshift to lower \Lssfr\ values at lower redshift. 

\editone{The evolution of the transitioning region is driven by the evolution of \lssfr\ as a function of redshift. At z $\sim$ 1.6 the population dominated by star-forming galaxies (\lssfr\ > -11.0) has \lssfr\ $\sim$ -9.7$^{+0.3}_{-0.7}$ while at z $\sim$ 0.85 it has \lssfr\ $\sim$ -10.0$^{+0.3}_{-0.4}$. A similar trend is seen in the population dominated by quiescent galaxies (\lssfr\ < -11.0) which has \lssfr\ $\sim$ -12.0$^{+0.7}_{-0.7}$ at z $\sim$ 1.6 and \lssfr\ $\sim$ -13.2$^{+0.8}_{-0.2}$ at z $\sim$ 0.85. This evolution in \lssfr\ for the star-forming dominated population is influenced by a $\sim$ 0.2 dex increase in stellar mass and $\sim$ 0.1 dex decrease in SFR over our redshift range with the evolution in the quiescent dominated population being solely driven by a 1.2 dex decrease in SFR as we see no change in stellar mass. \cite{bell04} states that the evolution they see in the blue cloud as a function of redshift (using colors) is caused by star-forming galaxies becoming more metal-enriched, dustier, and older at lower redshifts. The change we see in the blue cloud is in \lssfr, which is independent of metallicity or dust. Though we do see a correlation between the change in \lssfr\ and age. Therefore age may be the primary reason for the change in the transitioning region, with the older galaxies having lower SFRs and (in the case of the star-forming galaxies) higher stellar masses.}

\begin{figure*}[th]
\epsscale{1.2}
\plotone{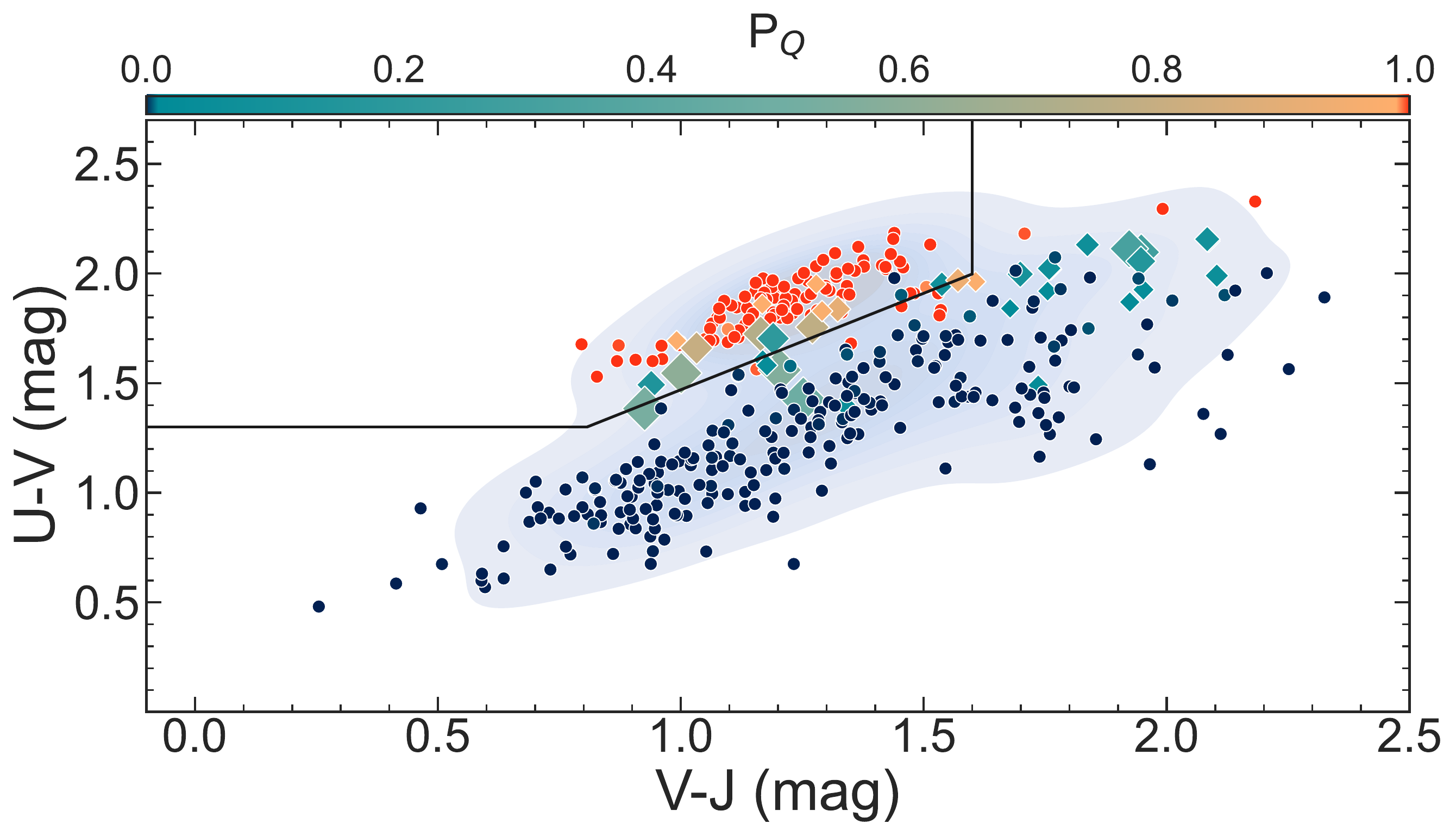}
\caption{A comparison of the UVJ rest-frame color-color diagram selection method to \pq\ \editone{with restframe colors derived using \texttt{eazy-py}}. Here the colors and shapes roughly correspond to red points/quiescent, green diamonds/transitioning, and blue points/star-forming, with the sizes of the points scaled by their distance from \psf\ $=$ 0.5 (where \psf\ $=$ 0.5 are the largest points) to emphasize the green valley. Here we see that the UVJ selection and \pq\ mostly agree with their classifications of star-forming and quiescent galaxies. As UVJ selection has no way of classifying green valley galaxies they tend to sit scattered around the quiescent limits (outlined polygon). Here we see the power in our method as we are not limited by discrete choices as \pq\ is a continuous classification.  
\label{fig_uvj}}
\end{figure*}

It is also essential to understand how our selection approach compares to the conventional methods. In Figure \ref{fig_uvj} we compare \pq\ to UVJ diagram selection. UVJ selection works as the diagram tracks how galaxies redden, going diagonally up to the right for dust reddening (dusty star-forming and quiescent galaxies lie in the top right corner) and going diagonally up to the left for reddening through aging. Galaxies are separated into quiescent and star-forming using the limits shown in Figure \ref{fig_uvj} defined in \cite{whit11}. We see that \pq\ mostly agrees with the UVJ selection with 92$\%$ of our quiescent classifications agreeing, 99$\%$ of our star-forming classifications agreeing, and 76$\%$ of our green valley galaxies being classified as star-forming. One takeaway from Figure \ref{fig_uvj} is that transitioning galaxies sit at the border of the selection criteria with large scatter, and therefore there is no easy way to use the UVJ diagram to specifically select transitioning galaxies. Indeed, this is one motivation for our approach to assigning a likelihood that a galaxy is transitioning. 
\subsection{Stacked Grism Spectra}

\begin{figure*}[th]
\epsscale{1.15}
\plotone{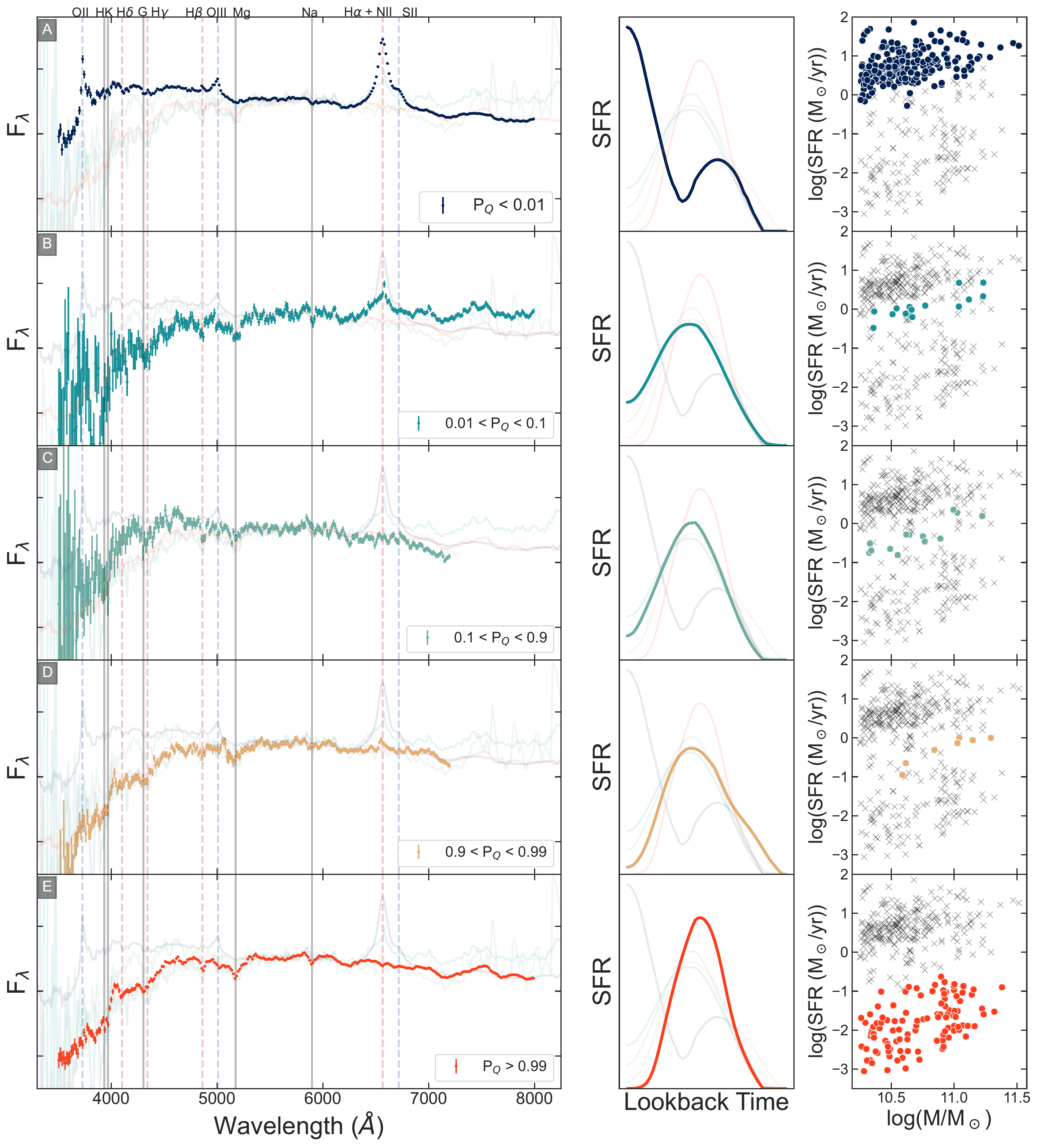}
\caption{Stacked spectra, stacked SFHs, and position on the star-forming main sequence for several different populations. \editone{In the stacked spectra/SFHs panels we include low opacity versions of all spectra/SFHs}. Panel \Panel{A}\ shows the star-forming galaxies with \psf\ $<$ 0.01 which have clear H$\alpha$, [OIII], H$\beta$, and [OII] emission. In Panel \Panel{B}\ we show galaxies with 0.01 $<$ \pq\ $<$ 0.1 which have H$\alpha$ in emission. We show the transitional galaxies (0.1 $<$ \psf\ $<$ 0.9) in Panel \Panel{C}\ which have possible H$\alpha$ emission and several absorption features. Panel \Panel{D}\ shows the late-phase transitional galaxies with 0.9 $<$ \psf\ $<$ 0.99, this spectrum has all the expected features of a quiescent galaxy, 4000 \AA\ break, Balmer absorption lines, several other metallicity absorption features (Ca HK, G, Mgb, Na), but has H$\alpha$ emission. In Panel \Panel{E}\ we show the spectra of the galaxies with the highest probability of being quiescent (\pq\ $>$ 0.99), we see all expected features of a quiescent galaxy with clear 4000 \AA\ break, Balmer absorption, and several metal absorption features. The central panels show the respective stacked star-formation histories, showing lower SFRs and older formation and we progress to higher \pq\ samples.
\label{fig_stk_samp1}}
\end{figure*} 

\begin{figure*}[th]
\epsscale{1.15}
\plotone{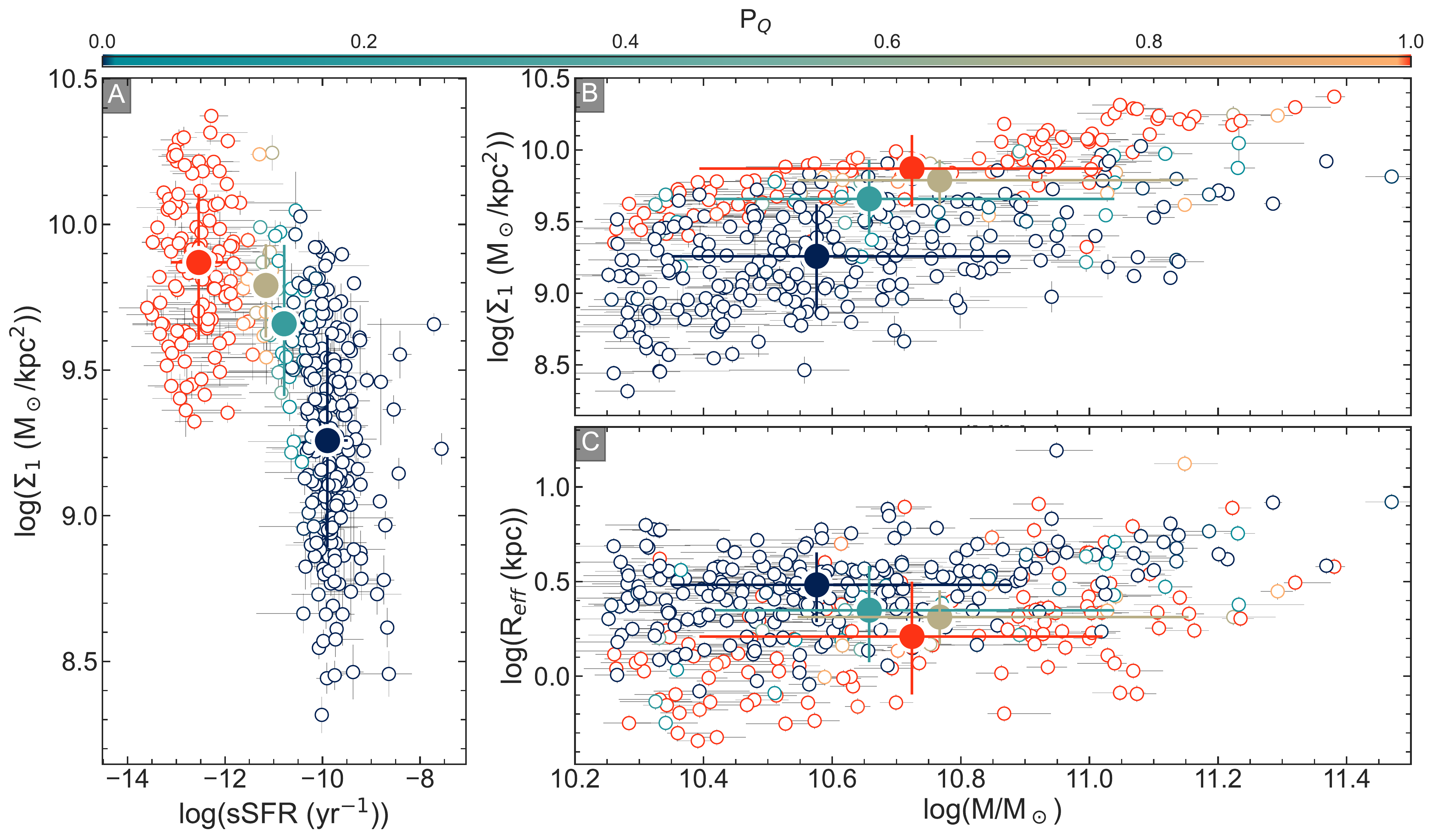}
\caption{Morphological relationships: \lssfr\ - \lsig\ in Panel \Panel{A}, stellar mass -  stellar-mass surface density in Panel \Panel{B}, and mass - size in Panel \Panel{C}. Individual galaxies are colored by their \pq\ values with larger filled points indicating the median values of the star-forming (blue), early green valley (cyan), late green valley (taupe), and quiescent (red) galaxies with error bars indicating the inner 68 percentiles. The median points create a crude evolutionary pathway that shows that galaxies with lower sSFRs / higher \pq\ values have more compact morphologies. \label{fig_mass_s1}}
\end{figure*} 

Using \pq\ we can see how the spectra and SFHs of galaxies evolve as they transition, we show this in Figures \ref{fig_stk_samp1} using stacked spectra and SFHs. Here we stack galaxies using a weighted mean (using two weighting terms).  The first weights are the measurement errors derived from the Grizli extractions, the second is a weight derived using a leave-one-out process \cite{estr19}. This works by taking a sample of spectra, stacking them using a weighted stacking (using the measurement errors only), then taking the same sample and, leaving one galaxy out, re-stacking. We then measure the variance between the two stacked spectra (the full stack and the stack with one galaxy left out). This variance is our second weight and will down-weight the effect outlier spectra have on the stack. We also normalize the grism spectra in the rest-frame region of 5400 $<$ $\lambda$ $<$ 5800 \AA\ using the best-fit FSPS models. 

Our SFHs are stacked using a bootstrapping method where we take the posteriors for the parameters needed to produce the Dense Basis SFHS (sSFR, $t_{25}$, $t_{50}$, $t_{70}$), sampling from them randomly producing 1000 SFHs all set at a redshift of 1. The resulting SFH is then the 50th percentile as a function of look-back time.

In Figure \ref{fig_stk_samp1} we examine how the spectra, SFH, and location on the star-forming main sequence evolve as galaxies quench.  Figure \ref{fig_stk_samp1} Panel \Panel{A} shows the \pq\ $<$ 0.01 galaxies that make up our star-forming sample. As we can see in the stacked spectra these galaxies show all the expected properties of star-forming galaxy spectra with several emission lines including H$\beta$, H$\alpha$, [OII], and [OIII] along with the absence of a 4000 \AA\ break. The SFH of this sample has a relatively high SFR (as the SFHs are normalized the actual SFR of the SFH is not meaningful). The last plot of Figure \ref{fig_stk_samp1} Panel \Panel{A} shows that these galaxies make up the blue cloud of the SFR - \lmass\ relationship.

 Figure \ref{fig_stk_samp1} Panel \Panel{B} shows the first of our transitioning spectra (0.01 $<$ \pq\ $<$ 0.1). Going from Panel \Panel{A} to \Panel{B} several emission lines weaken (with only H$\alpha$ surviving) along with a large change to the continuum at wavelengths below $\lambda$ $<$ 5500~\AA\ resulting in a possible Balmer-break along with several absorption features (H$\beta$, H$\gamma$, and Mg) strengthen. The SFH in Panel \Panel{B} shows that overall the SFH is dropping in this population, signaling that this population has already begun the quenching process with a SFR of 20$\%$ that of the star-forming sample. This population sits at the edge of the blue cloud, and therefore these are galaxies that have just entered the green valley.

The next Panel down (\Panel{C}) shows the transitioning galaxies with 0.1 $<$ \pq\ $<$ 0.9. The stacked spectra in \Panel{C} still shows no clear 4000 \AA\ break while still having possible H$\alpha$ emission along with several absorption features. The stacked SFH is similar to what is expected for a quiescent galaxy with a slightly higher SFR ($\sim$ 11$\%$ the SFR of the star-forming sample), They also sit in the middle of the green valley of the SFR - \lmass\ relationship. 

In Figure \ref{fig_stk_samp1} Panel \Panel{D} we see the transitioning galaxies with 0.9 $<$ \pq\ $<$ 0.99. The stacked spectra in \Panel{D} shows a clear 4000 \AA\ break as these galaxies seem to be essentially quiescent with what appears to be a small amount of H$\alpha$ emission. The SFH in \Panel{D} has an SFR of 3$\%$ the SFR of the star-forming sample and these galaxies sit at the edge of the red cloud. 

The galaxies which make up Figure \ref{fig_stk_samp1} Panel \Panel{E} are our quiescent sample, \pq\ $>$ 0.99. The stacked spectra show all the expected features of quiescent galaxies with a clear 4000 \AA\ break, several absorption features, and no emission. The SFH shows that most of the stellar mass formed long ago and the galaxy has been quenched for some time with an SFR of 0.04$\%$ the SFR of the star-forming sample. These galaxies make up the red cloud of the SFR - \lmass\ relationship.

\section{Discussion \label{sec_dis}}
With \pq\ we are able to view the evolution of properties as galaxies quench, we are also able to isolate quiescent, transitioning, and star-forming populations. Combining our flexible SFHs with \pq\ will also allow us to track the evolution of properties and link the evolution of common properties within groups.

\begin{figure*}[th]
\epsscale{1.15}
\plotone{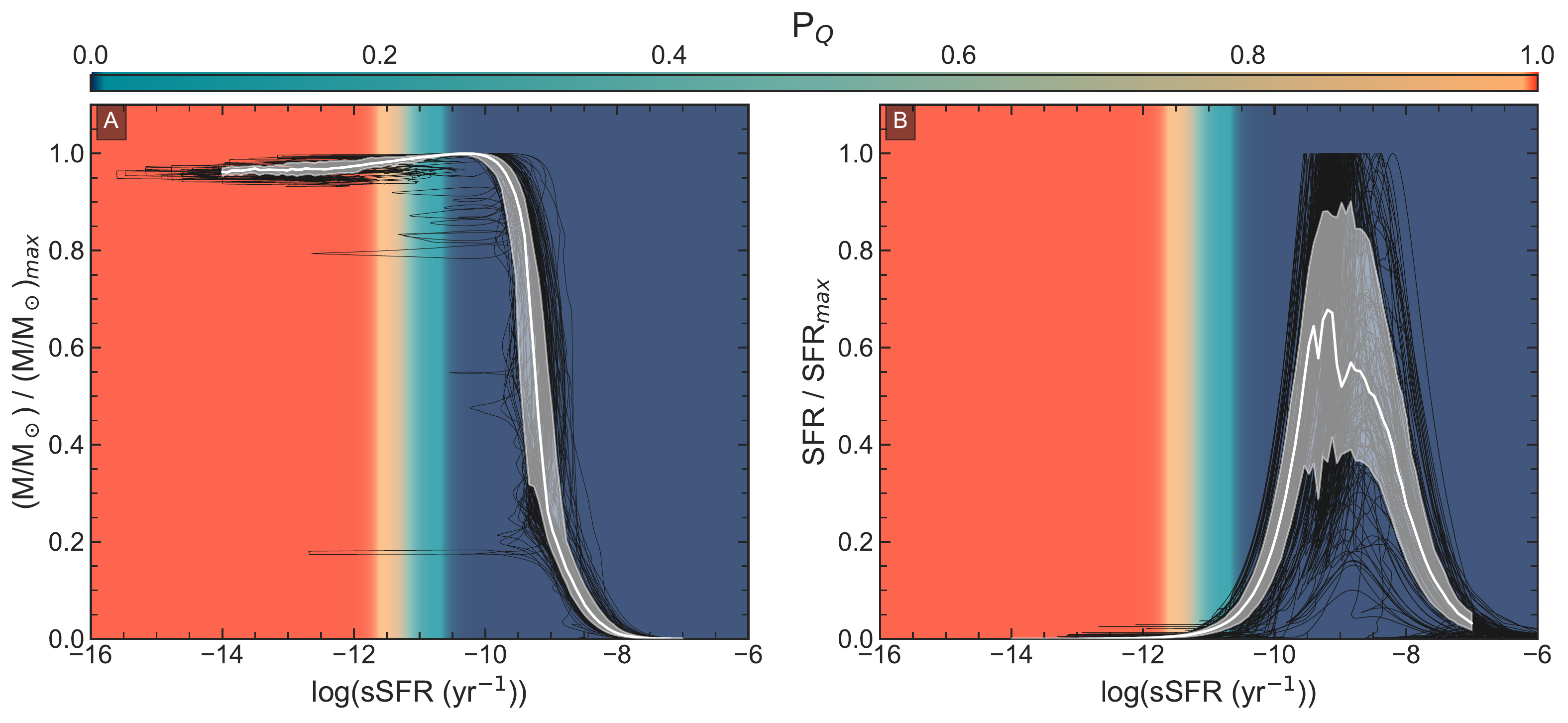}
\caption{SFH properties as a function of \lssfr. 
Panel \Panel{A} shows the mass formation percentage for our sample of quiescent galaxies (black lines) with the median trend shown as a white line and the inner 68th percentile shown as a grey region. Panel \Panel{B} shows SFR / SFR$_{peak}$ as a function of \lssfr\ for our sample of quiescent galaxies (\pq\ $>$ 0.99, black lines) with the median trend shown as a white and the inner 68th percentile shown as a grey region. Both panels are shaded to show \pq\ as a function of \lssfr\ at our median redshift. In Panel \Panel{B} we see that star-formation peaks in for most galaxies in a range of -9.5 $<$ \lssfr\ $<$ -8.5, after which star formation begins to drop well before the galaxy enters the green valley. We see that when entering the green valley mass formation is essentially done for most galaxies, with the median trend in Panel \Panel{A} showing that when galaxies enter the green valley they are at their most massive reaching their peak mass at \lssfr\ $\sim$ -10, then declining likely due to losses from stellar evolution. 
\label{fig_mevo}}
\end{figure*} 
\begin{figure*}[th]
\epsscale{1.15}
\plotone{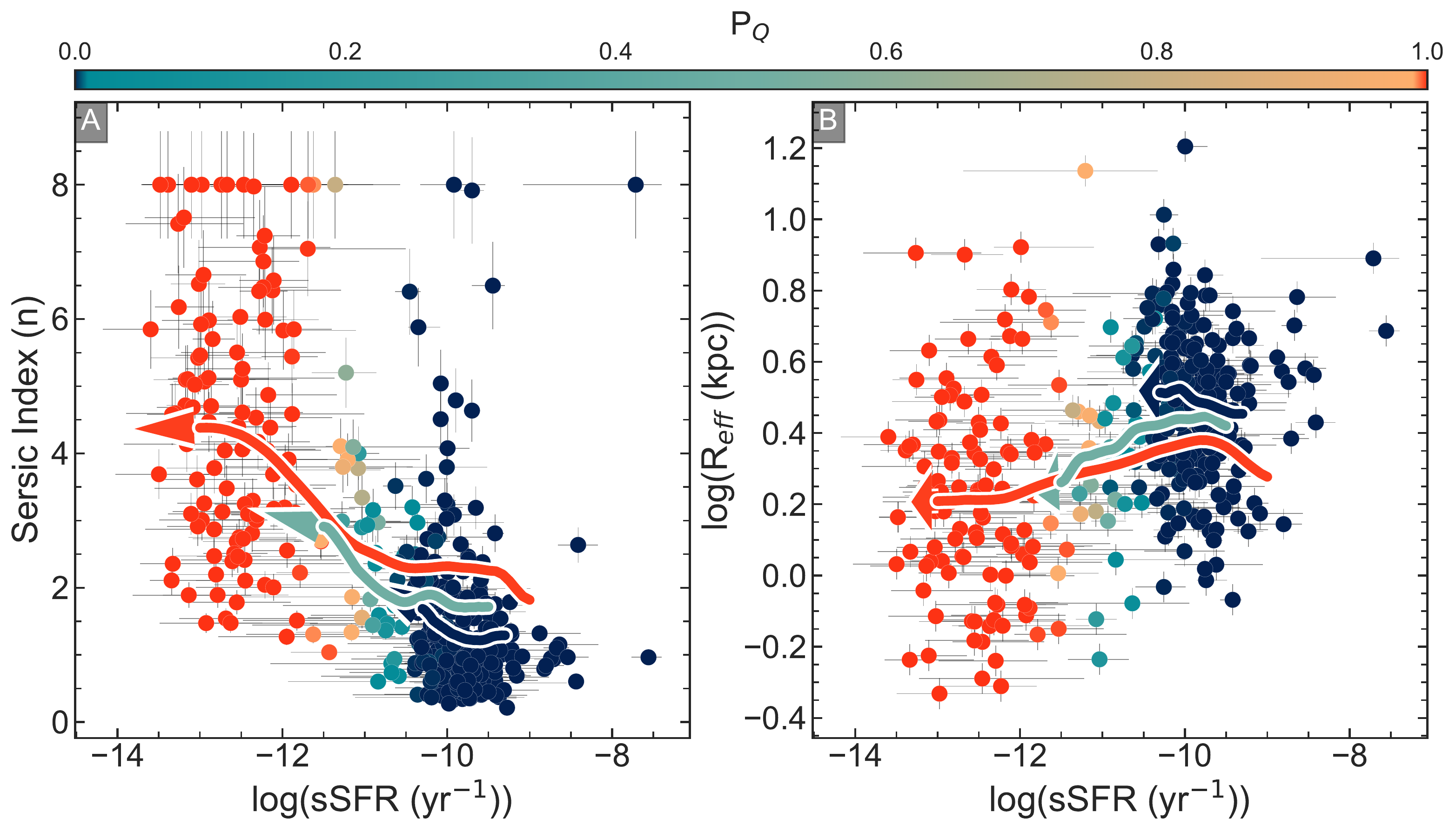}
\caption{Relationship between morphological properties \citep{vanw14} and \lssfr. Here we show individual galaxies colored by \pq\ with the median evolutionary pathways for star-forming (blue) - green valley (cyan) - quiescent (red) galaxies with Panel \Panel{A} focusing on Sersic index and Panel \Panel{B} showing $R_{eff}$, the arrows show the direction of the evolution of \lssfr. The evolution of these properties with \lssfr\ shows that as galaxies become quiescent they become bulge dominated and have smaller R$_{eff}$.
\label{fig_ssfrevo}}
\end{figure*} 
\begin{figure*}[th]
\epsscale{1.15}
\plotone{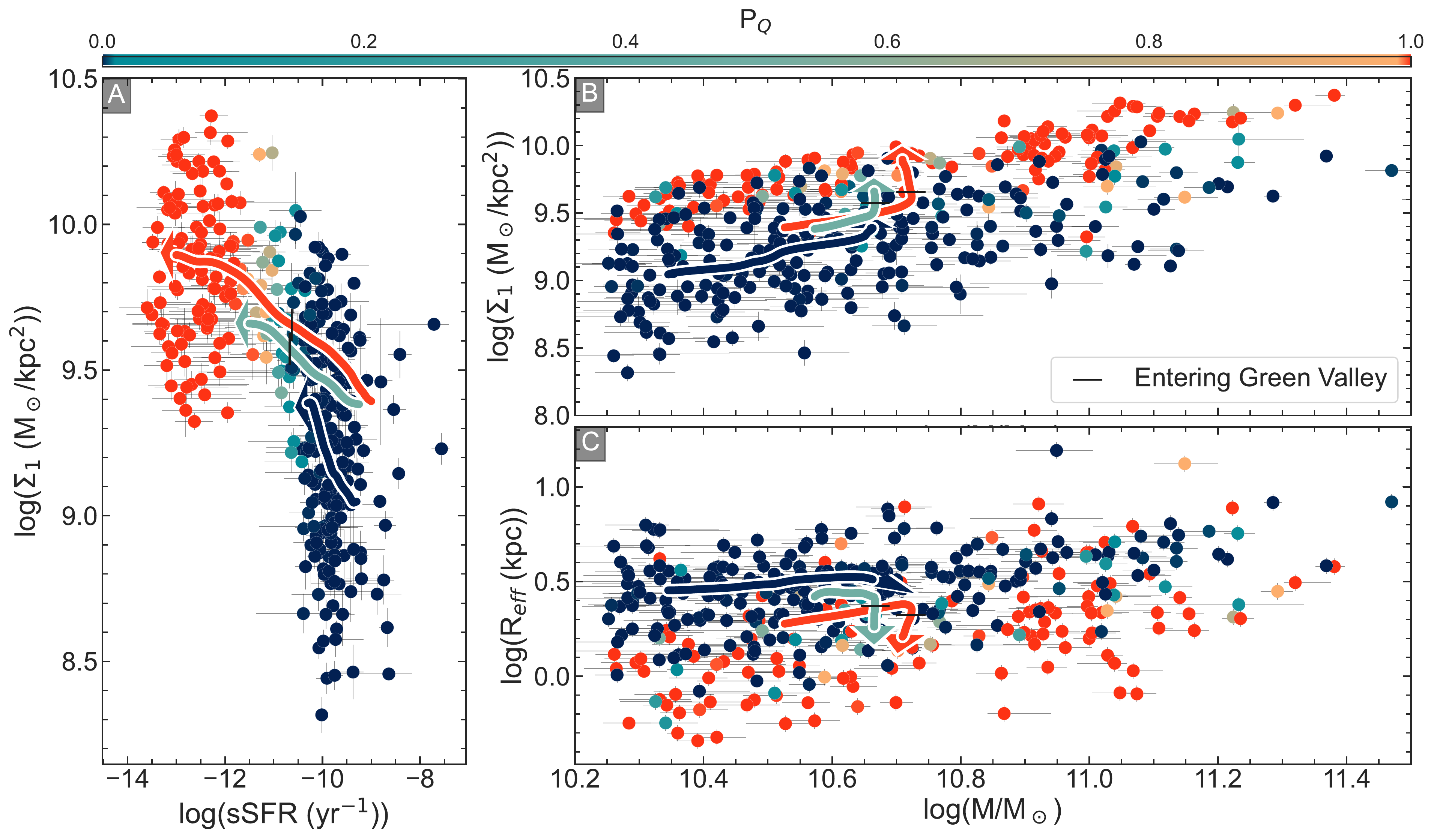}
\caption{Median evolutionary paths of our sample. In all panels, the points are colored by their \pq\ values with the curve indicating the median paths of our star-forming (blue) - green valley (cyan) - quiescent (red) galaxies. Panel \Panel{A} shows the evolution of \lsig\ with \lssfr\,  Panel \Panel{B} we see the stellar mass -  stellar-mass surface density relationship, and Panel \Panel{C} shows the mass - size relationship with arrows showing the direction of the evolution of \lssfr. These evolutionary paths show us that galaxies become compact while crossing the green valley and as only $\sim$ 2.4$\%$ of their mass is formed after the onset of quenching, the change in \lsig\ is purely morphological.
\label{fig_path}}
\end{figure*} 

Our goal for this work is to use \pq\ to classify galaxies in order to track how morphology evolves alongside quenching and to answer when do quiescent galaxies become more compact than star-forming galaxies. Several studies \citep{barr13,sues21} have shown an apparent change in morphology as a function of SFR and we can use the relationships seen in these works to determine the evolutionary path of morphology. Figure \ref{fig_mass_s1} shows several relationships between morphology and \lssfr\ / \lmass. Stellar-mass surface density \lsig\ is the parameterization of compactness we use. In Figure \ref{fig_mass_s1} individual galaxies are shown as outlines colored by their \pq\ values with large filled points indicating the median of the star-forming (blue - \pq\ $<$ 0.01), early green valley (cyan - 0.01 $<$ \pq\ $<$ 0.5), late green valley (taupe - 0.5 $<$ \pq\ $<$ 0.99), and quiescent (red - \pq\ $>$ 0.99) galaxies with error-bars indicating the inner 68 percentiles of the samples.

Panel \Panel{A} of Figure \ref{fig_mass_s1} shows the sSFR - $\Sigma_1$ relationship. Here we see that the blue cloud has a large range of \lsig\ values while the red cloud has a smaller range with higher overall \lsig\ values. The median points outline a crude trend that shows that galaxies with lower sSFRs/ higher \pq\ values have more compact morphologies, with the largest difference occurring between the star-forming and early green valley galaxies. A similar trend is seen in the stellar mass - stellar mass surface density relationship shown in Panel \Panel{B}, where we also see that the median stellar mass increases for galaxies with higher \pq\ values, with the late green valley galaxies sitting at higher mass than the quiescent galaxies though with only 13 galaxies making up this group this could simply be due to the small sample size. In Panel \Panel{B} we also see that quiescent galaxies are more compact than star-forming galaxies at fixed masses. Figure \ref{fig_mass_s1} Panel \Panel{C} shows the size-mass relationship, where we see that as we go from star-forming to quiescent sizes become smaller. The crude trends seen in Figure \ref{fig_mass_s1} suggest that after a galaxy ends its star-forming phase and enters the green valley it goes through a morphological change until it finally reaches the red cloud (galaxies may go through additional passive morphological evolution but that is beyond the scope of this project). Though, we should be cautious of using the trends seen here as it will suffer from the \editone{progenitor effect} \citep{ji22a,ji22} because we are drawing trends from star-forming to quiescent when the star-forming galaxies here are not the progenitors of our quiescent sample.

To investigate some of the effects of the \editone{progenitor effect} we will model the evolution of $\Sigma_1$ with respect to \lssfr\ (and therefore as a function of quenching) for our galaxies. $\Sigma_1$ is the stellar-mass surface density within 1 kpc and is defined as
\begin{equation}
\Sigma_1 = \frac{\int _0 ^{1~\mathrm{kpc}}\ I(R)\ 2 \pi R\ dR}{\int _0 ^{\infty}\ I(R)\ 2 \pi R\ dR} 
\frac{M_{*}}{\pi(1\ \mathrm{pkpc})^2}
\label{eq1}
\end{equation}
where $I(R)$ is the Sersic profile defined as 
\begin{equation}
I(R) = I_e \exp{\left(-b_n \left[\left(\frac{R}{R_{eff}}\right)^{(1/n)} -1\right]\right)}
\label{eq2}
\end{equation}
\citep{sers63, sima02}, where $R_{eff}$ is the effective radius, $I_e$ is the intensity at $R_{eff}$, $n$ is the Sersic index and $b_n$ is a coefficient dependent on n. Therefore to understand the evolution of compactness (\Lsig) with quenching we need to understand how stellar mass, $R_{eff}$, and $n$ evolve with \lssfr.

The evolution of stellar mass is detailed in our flexible SFHs. We can derive the stellar mass formed at any point in our look-back time using FSPS. FSPS will allow us to measure stellar mass formed as a function of look-back time while folding in the evolution of the stellar populations, most importantly the loss of stellar mass as a function of age. With the mass-formation history (MFH) we can derive the specific SFH (sSFH) which is the SFH divided by the MFH. As the sSFR is more related to the star-formation activity level of a galaxy (i.e. its \pq) than its SFR we will use the sSFH as a measure of the galaxy's star-formation activity. Figure \ref{fig_mevo} Panel \Panel{A} shows the relationship between the MFH / MFH$_{peak}$ and sSFH for our quiescent population (\pq\ $>$ 0.99) with individual evolutionary paths shown in black and the median path shown in white with the inner 68 percentile shown as a grey shaded region, the background is shaded by \pq\ (as \pq\ evolves with redshift this is only a median relationship with \lssfr). In Panel \Panel{A} we see that nearly all mass is formed between -10 $<$ \lssfr\ $<$ -8 and that galaxies are their most massive around when they enter the green valley. Figure \ref{fig_mevo} Panel \Panel{B} shows the relationship between SFH / SFH$_{peak}$ and sSFH with individual evolutionary paths shown in black and the median path shown in white with the inner 68 percentile shown as a white shaded region, the background is shaded by \pq. We see here that between -9.5 $<$ \lssfr\ $<$ -8.5 is when galaxies hit their peak SFRs and by the time they reach the green valley their SFRs are $\sim$ 7.5$_{-1.6}^{+5.1}$ $\%$ their peak SFRs and $\sim$ 0.2$_{-0.04}^{+0.13}$ $\%$ their peak SFRs when they are quiescent. We estimate that galaxies form $\sim$ 2.4$_{-0.4}^{+1.9}$ $\%$ of their stellar mass during the green valley, and as we see in Panel \Panel{A} during the green valley galaxies are losing stellar mass (to the ISM) faster than they are building it up. Therefore any change seen in \lsig\ during the green valley would not come from stellar mass and instead must come from either change to the Sersic index or $R_{eff}$.

\subsection{The Evolution of Morphology}

To model the evolution of $\Sigma_1$ we must first model the evolution of the Sersic index and $R_{eff}$, which we do by using MFHs and sSFHs using the following method.

\begin{itemize}
    \item Step 1) We trace the stellar mass as a function of redshift for each galaxy, finding galaxies in the \threedhst\ GOODS catalogs \citep{skel14} at similar masses at these redshifts out to z $=$ 3. We find that our stellar masses compare with the \threedhst\ with an offset of -0.04$_{-0.22}^{+0.17}$ dex, therefore we do not correct for systematic bias in the stellar masses and we search for galaxies in a 3$\sigma$ ($\sim$ 0.5 dex) range for stellar mass and a dz $\sim$ 0.5 for redshift.

    \item Step 2) We classify our list of progenitors from the GOODS catalogs as star-forming or quiescent using the UVJ diagram \citep{whit11} using the appropriate redshift limits. We then use the sSFHs to classify our galaxies as quiescent, green valley, or star-forming as a function of redshift using a linear fit to the \pq\ $=$ (0.01, 0.99) bounds out to a redshift of z $=$ 3. Using these classifications we can filter our list of progenitor galaxies to match stellar mass and star-formation activity levels. This allows us to take currently quiescent galaxies and compare them to star-forming galaxies when appropriate (determined by the sSFH) though as the UVJ diagram has no green valley classification when our galaxies are in the green valley we compare to both quiescent and star-forming galaxies.

    \item Step 3) Using the \cite{vanw14} GOODS morphological catalogs we then reconstruct the Sersic index and $R_{eff}$ histories of our galaxies out to z $=$ 3 by taking the medians at each redshift bin and scaling the track to the appropriate final Sersic index and $R_{eff}$ value, here we use the recommended filter with respect to redshift \cite{vanw14}. This scaling is done as the median Sersic index and $R_{eff}$ at the redshift when the galaxy is observed may not match the galaxy's actual Sersic index and $R_{eff}$, so we correct for the offset.
    
\end{itemize}

Figure \ref{fig_ssfrevo} outlines the evolutionary tracks of Sersic index and $R_{eff}$. In both panels, points are colored by their \pq\ values and the median trends for our quiescent (red) - green valley (cyan) - star-forming (blue) samples are shown outlined in white. When comparing the three different tracks in Panels \Panel{A}\ and \Panel{B}\ we see that the overall trend is that as galaxies quench (go towards lower sSFR) they gain higher Sersic index (n $>$ 2) and with smaller $R_{eff}$. We also see that our quiescent sample came from a star-forming sample with larger Sersic indices and smaller $R_{eff}$ (with respect to the green valley and star-forming populations), and therefore the star-forming progenitors of our quiescent sample were more compact than our current star-forming galaxies. The difference between these tracks can be interpreted as the \editone{progenitor effect}. Our green valley galaxies follow the same trend with star-forming progenitors that were more compact than our current star-forming galaxies but less compact than our quiescent galaxies' progenitors.  

With the trend lines from Figure \ref{fig_ssfrevo} and the median MFHs from Figure \ref{fig_mevo} Panel \Panel{A}, we can estimate the evolution  of $\Sigma_1$ as a function of \lssfr. Figure \ref{fig_path} Panel \Panel{A} shows how most galaxies may evolve in the \lssfr\ - \lsig\ relationship. The trends seen here are the median paths of the quiescent (red) - green valley (cyan) - star-forming (blue) samples. Here we see that the star-forming galaxies have \lsig\ values that are rising sharply likely due to the larger amounts of mass being formed at this stage, i.e. larger SFRs. Our quiescent and green-valley samples have a much shallower slope with vertical dashes showing approximately when these tracks enter the green valley. We see an overall increase in \lsig\ post entering the green valley with the quiescent sample increasing \lsig\ by 0.24 dex. 

Figure \ref{fig_path} Panel \Panel{B} shows the stellar mass - stellar mass surface density relationship. The star-forming track shows that stellar mass and $\Sigma_1$ evolve linearly before the galaxy quenches, and we see in the green valley and quiescent tracks that once the galaxy enters the green valley (indicated by the dashed line) that the galaxy continues to increase in \lsig\ but not in stellar mass. A similar trend is seen in Panel \Panel{C} (size - mass relationship) where the evolution is linear between stellar mass and $R_{eff}$, and once entering the green valley the sizes become smaller but the mass does not increase. 

With the increases seen in \lssfr\ and the lack of growth in \lmass\ post peak star formation, we see evidence that the difference in compactness between quiescent and star-forming galaxies is almost purely morphological, specifically $R_{eff}$ and Sersic index and not related to stellar mass. We see the effects of the \editone{progenitor effect} in the offsets between the quiescent - green valley - star-forming trends. What we can conclude is that the \editone{progenitor effect} does play a role in the difference in compactness between star-forming and quiescent galaxies, yet galaxies still do go through a morphological change as they quench, though the change is much less than what is suggested in Figure \ref{fig_mass_s1}. 

A question that arises then is whether the change in morphology is an actual structural change or whether this effect is more related to a process like disk fading \citep{math20}. In the disk fading model galaxies are comprised of a densely populated bulge and less dense disk, assuming inside-out growth \citep{beza09, naab09, nels16} the disk will be comprised of younger/brighter stellar populations, and as the galaxy quenches the brighter disk fades while the older denser bulge is relatively unchanged. This process will result in stellar light-density profiles with smaller sizes, and thus more compact quiescent population. Therefore these results may change if we used morphological results from stellar mass maps.  While this is beyond the scope of this project, it will be feasible with multiband high-angular resolution imaging now available with \hst\ and \jwst\ observations. 

\subsection{Our Results in Context - Bulge Growth vs Disk Fading}

\editone{Other works have examined the morphological evolution across the green valley, and by comparing to their findings we could begin to speculate as to what exactly is driving the evolution we see.} 

\editone{Our results favor disk fading being the dominant factor for the evolution we see in $\Sigma_1$. We see that the progenitor effect does impact the difference between the morphologies of star-forming and quiescent galaxies, though as seen in Figure \ref{fig_path} our evolutionary tracks still show an increase of $\Sigma_1$ when accounting for the progenitor effect. Additionally, we do not favor bulge growth as we see no increase in stellar mass as galaxies cross the green valley.}

\editone{In \cite{fang13} their green valley sample has similar $\Sigma_1$ values to their quiescent sample. These results suggest that the morphologies of quiescent galaxies are already set by the time the galaxy enters the green valley, which disfavors bulge growth. In \cite{brem18} it is found that green valley galaxies have significant bulge and disk components and that the morphological change from star-forming to quiescent is due to differences in their mass-to-light ratios. Disk fading is further supported by \cite{kim18} as they find that the presence of a bulge seems to be a prerequisite to quenching, and therefore the bulge may not need to grow during the green valley in order for the galaxy be bulge dominated when it quenches.  \cite{lope18} study the two-component nature (bulge and disk) of nearby galaxies. By measuring their spatially resolved SFHs they find that early-type spirals (likely the immediate progenitors of green valley galaxies) have older bulges that have little ongoing star formation while their disks are younger and have a more extended period of star formation. As our galaxies are field galaxies we do not expect their quenching to be driven by outside forces, therefore we would not expect the bulge to suddenly rejuvenate and build up mass again. The older bulges found in the early type spirals support disk fading as the mechanism for the change we see in $\Sigma_1$ as galaxies cross the green valley. Combing these results with our own, the story may be that star-forming galaxies form stars and eventually developing a bulge, they then begin the quenching process and enter the green valley. While in the green valley, the disk component fades and therefore the galaxy becomes bulge dominated by the time its quiescent.}

\editone{While our research favors disk fading, many other works have found evidence of bulge growth being the mechanism behind the evolution of morphology across the green valley. In \cite{guo21} they study the relationship between quenching and stellar mass surface density ($\Sigma_1$) for a sample of low-mass galaxies at 0.5 $<$ z $<$ 1.0, looking at the evolution of stellar mass surface density, stellar mass, SFR, and sSFR as galaxies cross the green-valley. They find that stellar mass increases ($\sim$ 0.1 - 0.2 dex) during the green valley phase whereas we find no increase in stellar mass. The difference is in large part due to our inclusion of stellar evolution in our MFHs, we find that during the green valley, galaxies form $\sim$ 2.4$^{+1.9}_{-0.4}\%$ of their stellar mass but are losing more mass than they gain due to stellar evolution. Additionally, \cite{guo21} finds that the long quenching timescales of low-mass galaxies ($\sim$ 4 Gyrs) are enough to explain the evolution in $\Sigma_1$, as it provides sufficient time for star formation in the green valley to increase $\Sigma_1$.}

\editone{\cite{wu20} examines galaxies with strong Balmer absorption to test the role of central starburst in quenching. A starburst could also explain why quiescent galaxies are more compact as there is a burst of star formation in the central region of the galaxy which would quench star formation and increase $\Sigma_1$ through bulge growth. They find that bluer quiescent galaxies in their sample have smaller R$_{eff}$. The interpretation here is that the bluer quiescent galaxies are more recently quenched than the redder quiescent galaxies. We see a similar trend with color and R$_{eff}$, but find no relationship between R$_{eff}$ and how recently a galaxy quenched. We suspect the disagreement is caused  by the selection of the strong Balmer absorption galaxies in \cite{wu20} which will preferentially select post-starburst galaxies whereas our sample has no such selection. The post-starburst sample will have fast quenching timescales while our sample is mostly made up of slower quenching galaxies.}

\editone{\cite{quil22} study the effects of both bulge growth and disk fading during the transition through the green valley for a sample of nearby galaxies. They find that the bugle-to-total ratio grows during the green valley, and that disk fading and bugle growth during this period are both likely responsible. They state that bulge growth in the green valley likely occurs due to mergers, and though not frequent, their estimated quenching timescale of 2-5 Gyrs would allow for enough mergers to occur. }

\editone{Bulge growth during the transition from star-forming to quiescent could increase $\Sigma_1$ as evidenced from \cite{guo21} and \cite{quil22} over long quenching timescales or \cite{wu20} for fast quenching post-starburst galaxies. Our galaxies are either too slow quenching for the starburst option or too fast quenching for star-formation or mergers to be the driving factor in the evolution of $\Sigma_1$. }

\section{Conclusion \label{sec_con}}

In this work, we use a novel approach to identify galaxies that are transitioning from star-forming to quiescent for a sample of massive galaxies (\Lmass\ $>$ 10.2) at 0.8 $<$ $z_{grism}$ $<$ 1.65 from the CLEAR survey using stellar population fits utilizing a forward modeling approach. Our approach to finding these transitioning galaxies uses the shape of the \Lssfr\ distribution to generate a probability if a galaxy is star-forming or quiescent, $P_Q$, where we define star-forming as, $\psf < 0.01$, quiescent as $\psf > 0.99$, and transitioning galaxies at $0.01 < \psf < 0.99$. We apply this method as a function of redshift to account for the evolution of the \Lssfr\ distribution. 

We find that the transitional phase for galaxies goes to lower \Lssfr\ values at lower redshifts (Figure \ref{fig_Psf}). This evolution is likely linked to the evolution of the star-forming main sequence \citep{whit12}. 

We compare \pq\ to the UVJ diagram (Figure \ref{fig_uvj}) to see how well they agree in their classifications \editone{using restframe colors derived from \texttt{eazy-py}}. We find good agreement between the two classifying schemes as with our quiescent sample being correctly classified in the UVJ diagram 92$\%$ of the time and 99$\%$ of star-forming galaxies falling into the star-forming region of the UVJ diagram. Of course, the UVJ diagram has no way of identifying green valley galaxies. We find that 76$\%$ of our green valley population is identified as star-forming in the UVJ diagram.

Using measured trends in the evolution of galaxy structural properties \cite{vanw14} and SFHs from our sample we interpret how the progenitors of galaxies in our sample evolve in redshift  with respect to stellar mass and morphology.  From those trends (Figures \ref{fig_ssfrevo} and \ref{fig_path}) we find the following:

\begin{itemize}
    \item At any given stellar mass the quiescent population is more compact than the star-forming population (Figure \ref{fig_mass_s1})
    
    \item From their mass formation histories we find that galaxies are their most massive at an \lssfr\ of $\sim$ -10, just prior to entering the green valley (Figure \ref{fig_mevo})
    
    \item After entering the green valley galaxies only form $\sim$ 2.4$^{+1.9}_{-0.4}\%$ of their stellar mass (Figure \ref{fig_mevo}), but lose more stellar mass than they gain due to stellar evolution
    
    \item The star-forming progenitors of our quiescent galaxies sample were more compact than our current star-forming sample (Figures \ref{fig_ssfrevo} and \ref{fig_path}))
    
    \item As for when do quiescent galaxies become compact it seems that this occurs after the galaxies leaves its peak star-formation period, once the SFH turns over and the galaxy begins the quenching process $\Sigma_1$ increases while stellar mass does not (Figure \ref{fig_path}))
    
    \item After entering the green valley galaxies increase \lsig\ by $\sim$ 0.24 dex while their stellar mass decreases by $\sim$ 0.01 dex, meaning the apparent compaction seen solely due to changes is in the Sersic profile ($R_{eff}$, $n$; Figure \ref{fig_path}))

\end{itemize}

A question we are left with is is the apparent morphological change seen an actual change to the galaxy's morphology or is it linked to a process like disk fading. \editone{By comparing to the literature we see that bulge growth during the green is correlated to how fast the galaxy crosses the green valley. Fast quenching galaxies can achieve bulge growth due to star-burst \cite{wu20}, and galaxies with extended quenching timescales of 2-5 Gyrs can achieve bugle growth due to mergers and star formation \cite{quil22, guo21}. Our quenching timescale of 0.7$^{+0.6}_{-0.2}$ Gyrs makes our sample too slow for starburst to be the driving mechanism, and too fast for mergers and star formation to make an impact on the bugle mass. Making disk fading the likely mechanism for the evolution we see in $\Sigma_1$.} Performing similar analysis by measuring the mass maps of these galaxies would allow us to better address this question and is a task left to spatially resolved studies.

\section*{Acknowledgements}
We thank our colleagues on the CLEAR team for their valuable conversations and contributions.  We are also grateful to the anonymous referee whose comments have improved the quality and clarity of this paper. VEC acknowledges support from the NASA Headquarters under the Future Investigators in NASA Earth and Space Science and Technology (FINESST) award 19-ASTRO19-0122.  This work is based on data obtained from the Hubble Space Telescope through program number GO-14227.  Support for Program number GO-14227 was provided by NASA through a grant from the Space Telescope Science Institute, which is operated by the Association of Universities for Research in Astronomy, Incorporated, under NASA contract NAS5-26555.  This work is supported in part by the National Science Foundation through grants AST 1614668. The authors acknowledge the Texas A\&M University Brazos HPC cluster and Texas A\&M High Performance Research Computing Resources (HPRC, \url{http://hprc.tamu.edu}) that contributed to the research reported here.

\software{Astropy \citep{2018AJ....156..123A, 2013A&A...558A..33A}, Matplotlib \citep{hunt07}, NumPy \citep{harris2020array}, SciPy \citep{virt20}, grizli \citep{grizli}, eazy-py \citep{bram08, eazy-py}, FSPS \cite{conr10}, Dense Basis \citep{iyer17}, Seaborn \cite{wask21}, Pandas \citep{reba22}, Dynesty \citep{spea19}}

\bibliography{main}{}

\end{document}